\documentclass[aps,pra,reprint,amsmath,amssymb,showpacs,groupedaddress,floatfix]{revtex4-1}
\usepackage{mathtools,graphicx,bbold,bm,color}
\usepackage[colorlinks,citecolor=blue,urlcolor=blue,linkcolor=blue,hyperindex,breaklinks]{hyperref}
\usepackage{ulem}

\definecolor{astral}{RGB}{46,116,181}
\definecolor{azulclaro}{RGB}{32,119,199}
\definecolor{cinves}{RGB}{18,144,129}
\definecolor{guinda}{RGB}{128,23,14}
\newcommand{\dd}{\text{d}}

\providecommand{\ket}[1]{\lvert #1 \rangle}

\begin{document}

\title{Hawking temperature in dispersive media: Analytical and numerical study}
\author{Alhan Moreno-Ruiz, David Bermudez}
\email{dbermudez@fis.cinvestav.mx}
\homepage{\\ https://www.fis.cinvestav.mx/{\raise.17ex\hbox{$\scriptstyle\sim$}}dbermudez/}
\affiliation{Departamento de F\'isica, Cinvestav, A.P. 14-740, 07000 Ciudad de M\'exico, Mexico}


\begin{abstract}
In the context of analog gravity the Hawking effect can be generalized to domains outside astrophysics. Arguably, the most successful systems for this analogy have been so far the sonic and the optical ones. However, problems arise in the analog systems as their dispersive effects are too large to be ignored, and this in turn modifies the usual thermal spectrum of Hawking radiation. In this work we perform analytical and numerical studies on how the velocity profile modifies the Hawking temperature in dispersive media, including some with direct experimental application.\\

PhySH: quantum aspects of black holes, laboratory studies of gravity, Bose-Einstein condensates, optical fibers, Kerr effect.
\end{abstract}

\maketitle

\section{Introduction}
K. Schwarzschild found the first exact solution to the inhomogeneous Einstein equations in general relativity \cite{Schwarzschild1916}, this solution hinted on a mysterious object from which nothing can escape, not even light. These objects were eventually called black holes. However, it was later realized by S. Hawking \cite{Hawking1974,Hawking1975} that if we consider a quantum field on top of the classical background found by Schwarzschild, thermal radiation is produced and escapes from the black hole to infinity. This is the Hawking radiation.

This prediction from forty five years ago is used to test our present candidates for quantum theories of gravity. Nevertheless, Hawking radiation rests on some dubious assumptions \cite{Helfer2003} and itself has not passed the ultimate test, i.e., experimental verification. One way out of this conundrum was proposed by W. Unruh \cite{Unruh1981} and it consists in using analog systems to learn more about the Hawking effect. Eventually, this proposal gave birth to a new area of research known as analog gravity.

In analog gravity, phenomena usually related to gravity are studied in other systems. The clearest example of this is Hawking radiation itself. Around the event horizon of an astrophysical black hole, the Hawking process leads to the creation of thermal radiation of quantum origin. In analog systems a process with similar kinetics leads to the creation of thermal radiation, that can be of classical (stimulated) \cite{Drori2019} or quantum (spontaneous) \cite{deNova2019} origin. Common analog systems include water tanks \cite{Rousseaux2008}, Bose-Einstein condensates \cite{Barcelo2001cqg,Steinhauer2016}, optical fibers \cite{Philbin2008,Bermudez2016pra}, microcavity polaritons \cite{Nguyen2015}, superconducting circuits \cite{Nation2012}, among others.

The study of analog gravity has already given insights on the unclear details of Hawking's derivation \cite{Visser2003}, such as the role of high-frequency modes, known as the trans-Planckian problem \cite{Brout1995}. Moreover, its development has also given new points of view in the theories of the analog systems. For example, the horizon physics has predicted new nonlinear effects in optical fibers that have been measured in the search of analog Hawking radiation: negative-frequency resonant radiation \cite{Rubino2012prl} and stimulated Hawking radiation of positive \cite{Philbin2008} and negative \cite{Drori2019} frequencies.

On the other hand, the derivation of Hawking radiation in the analog systems do not suffer from the high-frequency divergences that may occur in the astrophysical case. This is due to the role of dispersion. A frequency shift of any signal is followed by a shift on its group velocity that ultimately leads to a finite frequency shift, unlike the astrophysical case, where frequency shifts are infinite. Based on previous work by S. Corley \cite{Corley1998}, U. Leonhardt and S. Robertson \cite{Leonhardt2012} developed an analytical theory to obtain the effective spectrum of Hawking radiation in dispersive media. This equation considers only the topological nature of the solution of the dispersion relation in the complex plane and it should be related to new approaches by F. Biancalana's group \cite{Robson2019,Robson2019a}.

In this paper our goal is three-fold. First, we expand the analytical theory from the sonic to the optical case and we point out the differences in their derivation. Second, we apply this theory to new velocity profiles, solve the integrals using Cauchy's reside theorem, and obtain their analytical Hawking spectrum in terms an effective temperature. And third, we test our analytical formulas with numerical solutions and found an excellent agreement. Furthermore, we applied a second-order approximation to the commonly used velocity profiles tanh and sech${}^2$ and obtain the spectrum analytically (as an infinite series) and numerically.

\section{The analogs of the event horizon} \label{analog}

\subsection{Schwarzschild spacetime}
\begin{figure}
	\centering
	\includegraphics[width=85mm]{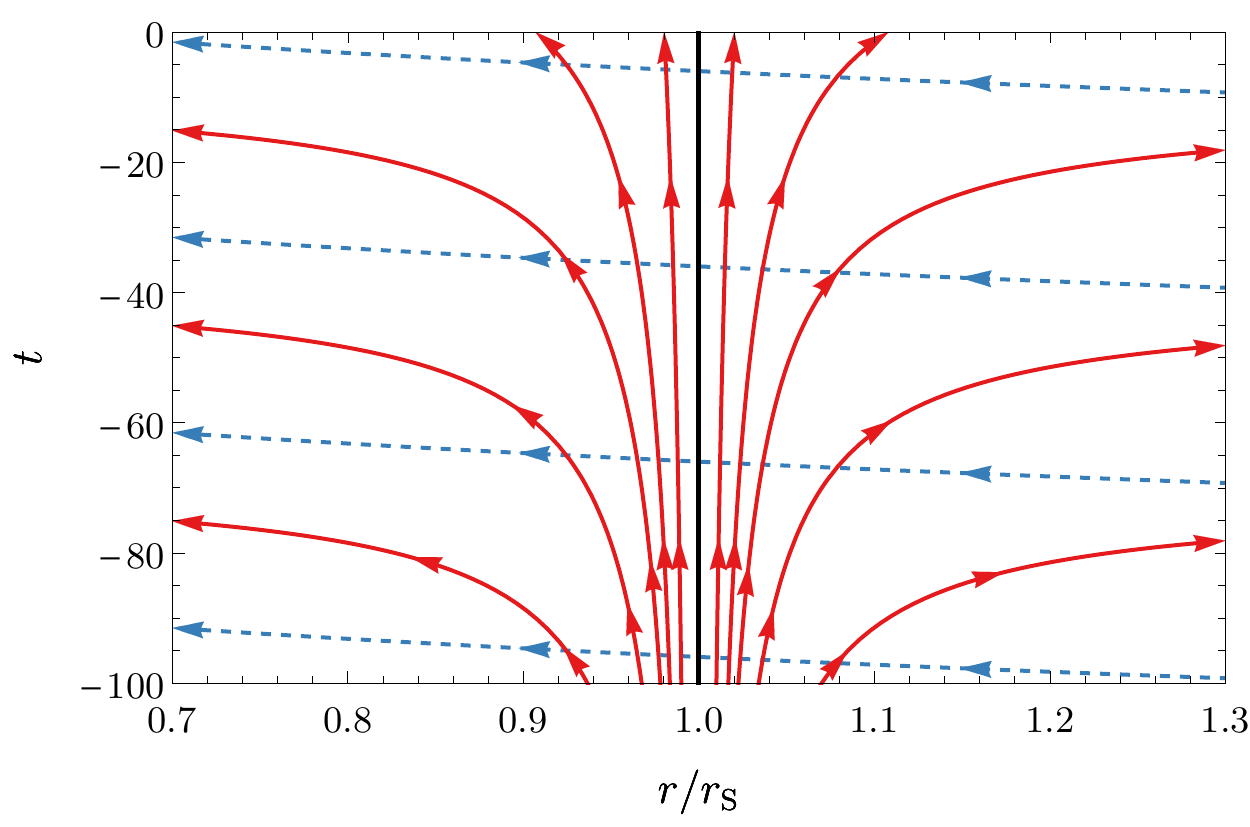}
	\caption{(Color online). Geodesics of light rays around the event horizon $r=r_\text{S}$ of a Schwarzschild spacetime. Counterpropagating u-modes are shown in solid red and copropagating v-modes in dashed blue.}
	\label{geodesics}
\end{figure}

To obtain an analog of the event horizon of a black hole, we consider the movement of light rays around the Schwarzschild radius $r_\text{S}=2GM/c^2$. This is given by the 1+1 Schwarzschild metric in Painlev\'e, Gullstrand, Lema\^itre coordinates \cite{Hartle2003}:

\begin{equation}
\dd s^2 = c^2 \dd t^2-\left(\dd r +\sqrt{\frac{r_\text{S}}{r}}c\dd t\right)^2.
\label{Sch}
\end{equation}
In Fig. \ref{geodesics} we show the light ray trajectories ($\dd s^2=0$) for outgoing (solid red curves) and ingoing (dashed blue lines) modes. Observe that ingoing modes do not have any peculiar kinematics at $r_\text{S}$, however, outgoing modes do, i.e., modes at $r<r_\text{S}$ cannot escape from the black hole geometry. These outgoing modes are the main interest for the derivation of Hawking radiation.

We can obtain the same kinematics by considering the black-hole spacetime as a moving medium, i.e., as a fluid whose movement is caused by gravity. In the sonic analog we consider sound waves in a moving fluid and in the optical analog, light waves propagating inside a dielectric material. Next, we will summarize and compare both analogies.

\subsection{Spacetime as a moving fluid}\label{stasfluid}
As we said, we can understand the spacetime in Eq. \eqref{Sch} as a fluid for outgoing waves. Then, the Schwarzschild spacetime flows towards the singularity with a varying velocity profile in the radial coordinate given by $c\sqrt{r_\text{S}/r}$ such that at the horizon ($r=r_\text{S}$) the spacetime velocity surpasses the speed of light, therefore trapping all light rays inside. Furthermore, we can generalize this metric to a general 1+1 velocity profile $u(z)$, as long as it surpass the limiting speed of the system $c$ (not necessarily the speed of light), this point will become the analog of the event horizon. The effective metric is then
\begin{equation}
	\dd s^2 = c^2 \dd t^2-\left(\dd z +u(z)\dd t\right)^2.
	\label{dsanalog}
\end{equation}
In the following, we assume that the moving medium moves to the left such that $u<0$.

\subsection{Sonic analog}\label{sonicanalog}
Let us consider a scalar massless acoustic field in 1+1 dimensions on top a moving fluid with metric \eqref{dsanalog}, i.e., the velocity of the medium varies with position $u=u(z)$. Then, under certain conditions \cite{Birrell1982} its Lagrangian can be written as
\begin{equation}
L=\frac{1}{2}\sqrt{-g}g^{\mu\nu}\partial_\mu\phi^* \partial_\nu \phi,
\end{equation}
where $g^{\mu\nu}$ is the inverse metric tensor and $g$ the determinant of the metric tensor $g_{\mu\nu}$. Using the metric \eqref{dsanalog} we obtain the wave equation
\begin{equation}
(\partial_t+\partial_z u)(\partial_t+u\partial_z)\phi-c^2\partial_z^2 \phi =0.
\end{equation}
We can generalize this equation by including the Bogoliubov dispersion term \cite{Leonhardt2007}, in that case the phase velocity is a function of the wavevector $c=c(k)$. As the medium moves to the left ($u<0$), the counterpropagating waves move to the right and they are the ones that can experience a horizon (the solid red lines in Fig. \ref{geodesics}). The dispersion relation of this medium is
\begin{equation}
\omega-u(z)k=c(k)k,
\label{disprel}
\end{equation}
and the Bogoliubov dispersion is
\begin{equation}
c(k)=c_0\sqrt{1-\frac{k^2}{k_0^2}},
\end{equation}
where the minus sign correspond to a supersonic dispersion. As long as the fluid velocity surpasses the speed of sound of the system $c_0$, a horizon can exists for certain frequencies.

In this work we will consider several velocity profiles $u(z)$. For example, in Eq. \eqref{Sch} the velocity profile can be taken as
\begin{equation}
u(z)=c\sqrt{\frac{r_\text{S}}{z}},
\end{equation}
Another example closer to the experimental implementation of sonic analogs \cite{Leonhardt2012} is a tanh profile
\begin{equation}
u(z)=\frac{u_2+u_1}{2}+\left(\frac{u_2-u_1}{2}\right)\tanh\left(\frac{z}{a}\right),
\end{equation}
where $u_2=-0.8$ and $u_1=-1.2$ are the subsonic and supersonic asymptotic velocities and $a=1/k_0$ determines the length scale of the velocity profile.

\begin{figure}
	\centering
	\includegraphics[width=85mm]{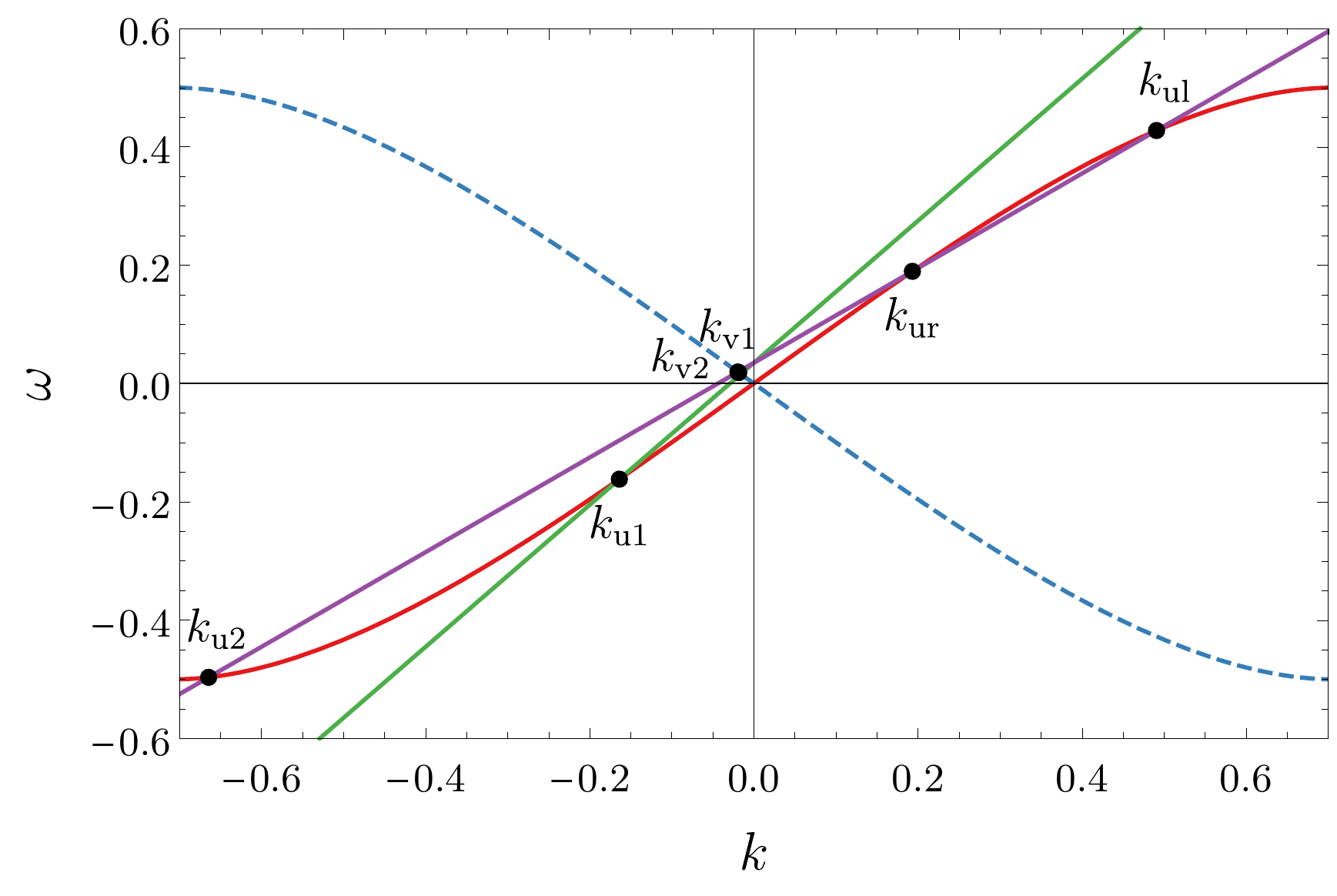}\\
	\includegraphics[width=85mm]{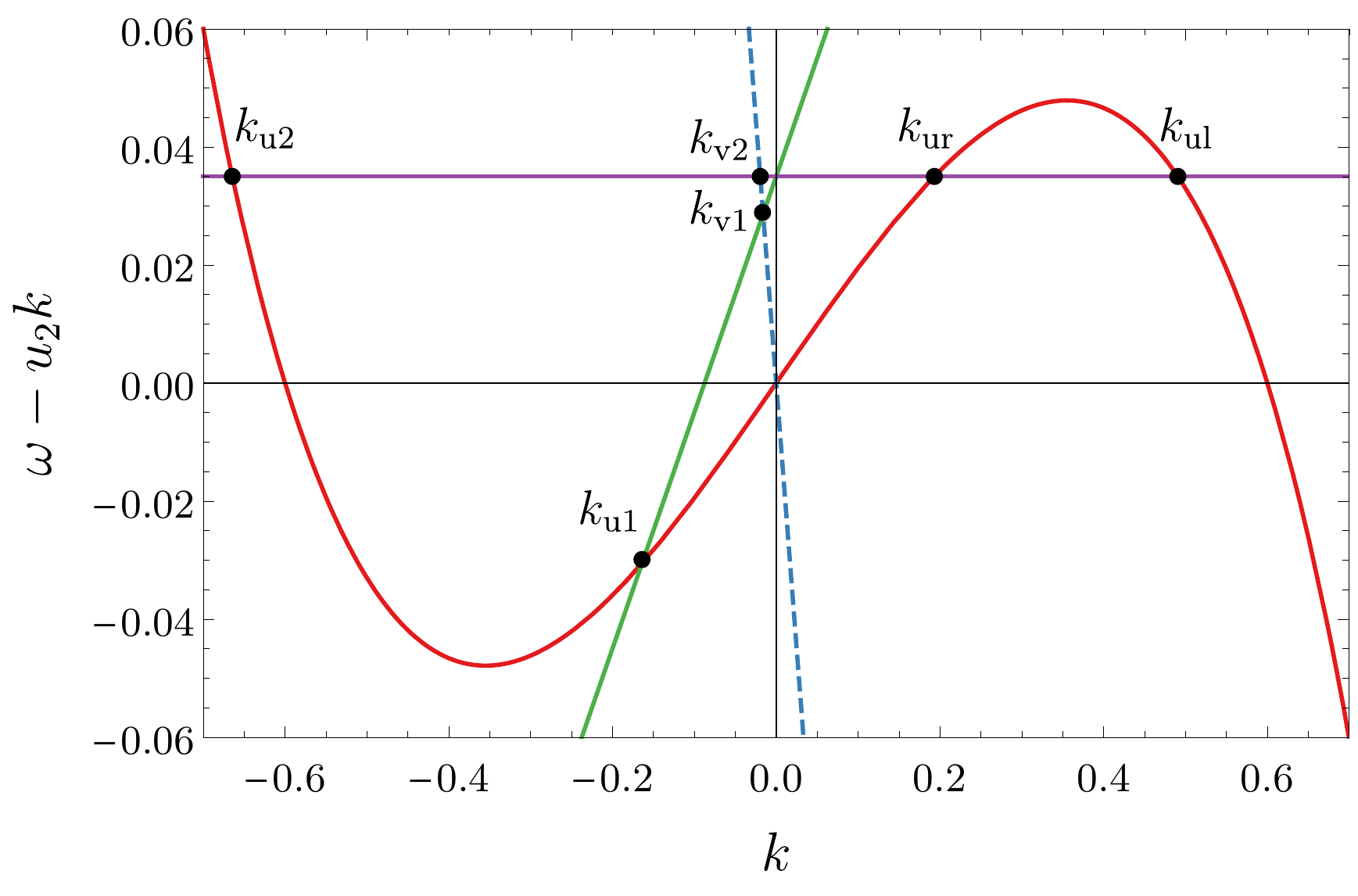}
	\caption{(Color online). Dispersion relations for the sonic case in the laboratory frame (top) and in the comoving frame (bottom). The counterpropagating u-modes (red) and copropagating v-modes (blue dashed) are matched to two straight lines with the two asymptotic velocities, one subsonic $u_2$ (purple) and one supersonic $u_1$ (green).}
	\label{dispsonic}
\end{figure}

We can obtain the asymptotic mode solutions for the dispersion relation as shown in Fig. \ref{dispsonic}. The counterpropagating u-modes are the ones suffering the horizon, for completeness we also show the copropagating v-modes.  The group velocity is $v_g=\partial_k\omega(k)=\partial_k[kc(k)]$, i.e., the slope of the curve $\omega(k)$, e.g., the solid red and dashed blue curves in Fig. \ref{dispsonic}(top). Modes with $v_g>|u|$ propagate to the right (such as $u_\text{ur}$) and modes with $v_g<|u|$ propagate to the left (such as $u_\text{ul}$). Further solutions $k_\text{u1}$, $k_\text{u2}$ propagate to the left and have negative wavenumber \cite{Biancalana2012}. It is also convenient to study the kinematics from a comoving frame with a given velocity $u_2$, the dispersion relation in this comoving frame is shown in Fig. \ref{dispsonic}(bottom).

If we consider wavepackets with central position $z$, wavenumber $k$, and frequency $\omega$, their dynamics is given by the Hamilton equations \cite{Leonhardt2007}:
\begin{equation}
\frac{\dd z}{\dd t}=\partial_k \omega=v_g+u,\quad
\frac{\dd k}{\dd t}=-\partial_z \omega=-k\partial_z u.\label{hamsonic}
\end{equation}
The group-velocity horizon is given by the condition $|u|=v_g$.

The phase is transformed as
\begin{equation}
\varphi=\int(k \dd z -\omega\dd t),
\label{phaseson}
\end{equation}
this is the essential quantity to achieve the phase-matching conditions. In other systems we should study how the phase is expressed, e.g., now we study the optical system.

\subsection{Optical analog}\label{opticalanalog}
In order to obtain an optical analog we would need a fluid moving close to the speed of light in media, which seems hard to achieve \cite{Leonhardt2002nat}. However, U. Leonhardt proposed a radical change of mindset \cite{Philbin2008}. Instead of a fast-moving fluid, we can send a refractive index perturbation using light pulses and study the system in its comoving frame. The change of frame ends up giving kinematic equations of the same form as those of the astrophysical case.

Let $n_0$ be the original refractive index of the material, a strong light pulse will change it by the optical Kerr effect \cite{Agrawal2013} as $n=n_0+\delta n$. We consider that this strong light pulse moves at a constant velocity $v_{g0}$, for example using fundamental solitons or other pulses whose dynamics is slower than the horizon physics \cite{Bermudez2016jp}. Then, in the comoving frame waves traveling with $v<v_{g0}$ move to one direction and those with $v>v_{g0}$ to the other direction (right and left, respectively, if we use the delay time $\tau$ in Eq. \eqref{tauzeta}).

Here we also model the dispersion properties of light inside a fiber with the wavenumber (usually denoted $\beta$ for fibers) given by
\begin{equation}
\beta(\omega)=\frac{\omega}{c}n(\omega).
\end{equation}
The simplest dispersion model that contains all the horizon physics \cite{Bermudez2016pra} is
\begin{equation}
\beta(\omega)=\frac{\omega}{c}\sqrt{1+\frac{\omega^2}{\omega_0^2}},
\end{equation}
where the sign of the second term means we refer to a subluminal ($+$) dispersion. Notice that this is the opposite sign as the one considered in the sonic case.

The comoving frame is defined by its velocity. In our case we always choose it as the group velocity of the Kerr perturbation $v_{g0}$. Then, the propagation time $\zeta$, that plays the role of time, and the delay $\tau$, that plays the role of distance, are
\begin{equation}
\tau=t-\frac{z}{v_{g0}},\qquad \zeta=\frac{z}{v_{g0}}.\label{tauzeta}
\end{equation}

\begin{figure}
	\includegraphics[width=85mm]{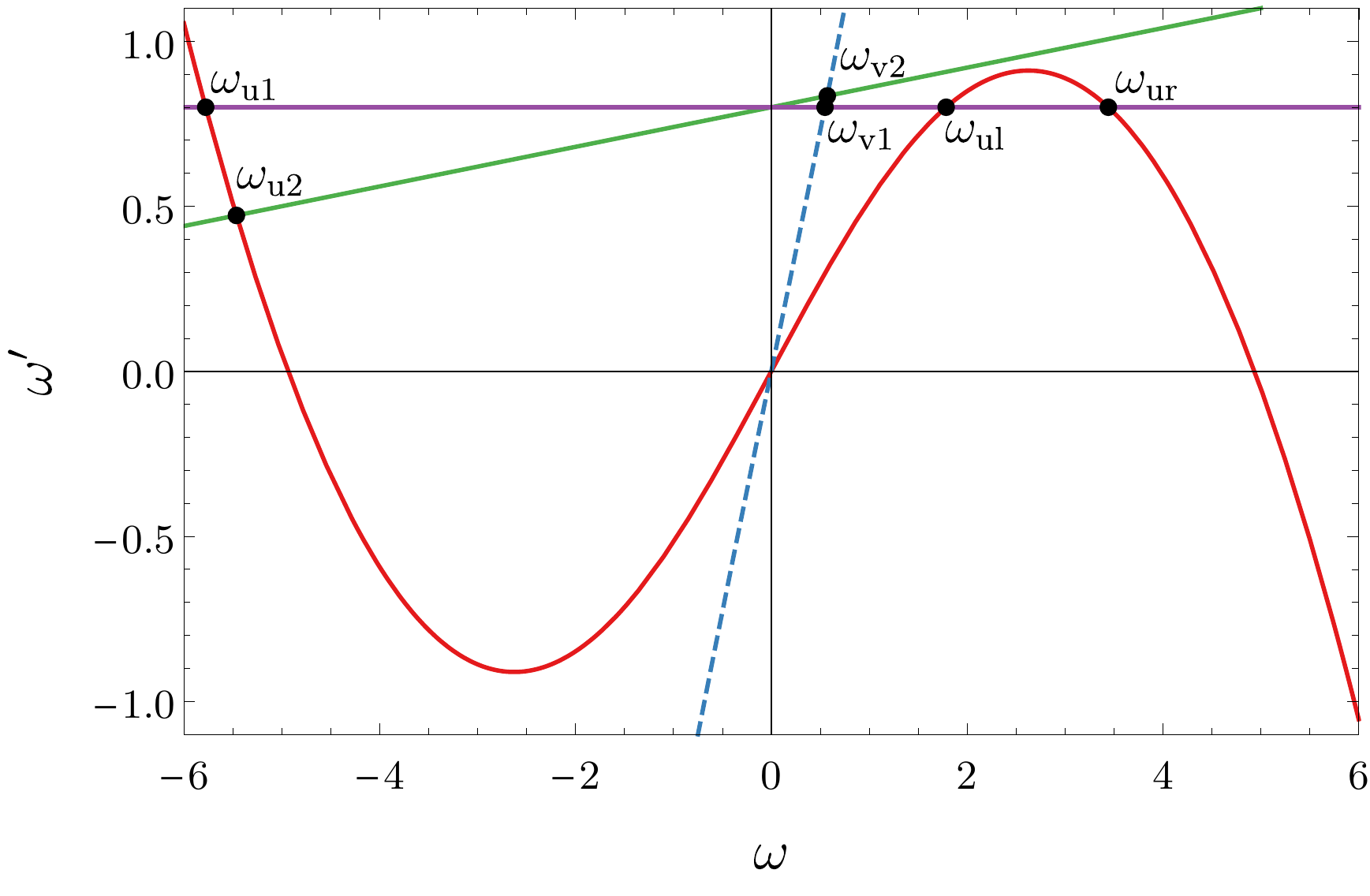}
	\caption{(Color online). Dispersion relation for the optical case in the comoving frame. Similar notation as in Fig. \ref{dispsonic} for the sonic case.}
	\label{dispopt}
\end{figure}

The phase is transformed from the sonic case \eqref{phaseson} to
\begin{equation}
\varphi=-\int(\omega\dd\tau+\omega' \dd\zeta ),
\label{phaseopt}
\end{equation}
where we have defined the comoving frequency $\omega'$ as
\begin{equation}
\omega'=\omega\left(1-\frac{n(\omega)}{n_{g0}}\right),
\end{equation}
with $n_{g0}=c/v_{g0}$. As can be seen by comparing Eqs. \eqref{phaseson} and \eqref{phaseopt}, the role of $k$ is now given to $-\omega$ and the role of $\omega$ to $\omega'$. The dispersion relation in the comoving frame in the optical case looks similar to that of the sonic case and it is shown in Fig. \ref{dispopt}. To complete the analogy with the sonic case, we should define the dimensionless quantity $v(\tau)$ that acts as the velocity profile for the sonic case
\begin{equation}
v(\tau)=n_{g0}-\delta n(\tau).
\label{vopt}
\end{equation}
As expected, $v(\tau)$ depends on the change due to the Kerr effect $\delta n(\tau)$. A comparison of the role that variables play in the sonic and optical systems is shown in Table \ref{table:so}.

\begin{table}
	\centering
	\begin{tabular}{|c | c|}
		\hline
		Sonic & Optical \\
		\hline
		$t$ & $\zeta$ \\
		$z$ & $\tau$ \\ 
		$k$ & $-\omega$ \\
		$\omega$ & $\omega'$ \\ 
		$c(k)$ & $n(\omega)$ \\
		$u(z)$ & $v(\tau)$ \\
		\hline
	\end{tabular}
	\caption{Relationship between the sonic and optical systems for analog Hawking radiation.}
	\label{table:so}
\end{table}

The Hamilton equations in the comoving frame are
\begin{equation}
\frac{\mathrm{d}\tau}{\mathrm{d}\zeta}=\frac{\omega'}{\omega}- \frac{\omega}{n_{g0}}\partial_\omega n(\omega), \quad \frac{\mathrm{d} \omega '}{\mathrm{d} \zeta}=-\frac{1}{n_{g0}}\partial_\tau\delta n(\tau).\label{hamoptic}
\end{equation}

In this case, the perturbation $\delta n$ is given by the pulse shape in $\tau$ and it defines the velocity profile $v(\tau)$, e.g., for optical systems the most used form is a soliton
\begin{equation}
\delta n(\tau)=\delta n_\text{max}\,\text{sech}^2\left(\frac{\tau}{a}\right),
\end{equation}
in any case, the shape of this perturbation should be localized and fixed in $\tau$, be it solitonic, Gaussian, super-Gaussian, or any other.

\section{Analytical theory}
The quantum vacuum state $\ket{0}$ is defined by the annihilation operator $\hat{a}$ through $\hat{a}\ket{0}=0$, $\forall\, \hat{a}$. However, in quantum field theory in curved spaces, the annihilation operator and the vacuum do not have a single definition in all space. This opens the possibility of creation of particles from vacuum. Some studied effects are the Hawking effect \cite{Hawking1974}, the Unruh effect \cite{Unruh1976}, the Casimir force \cite{Casimir1948}, the dynamical Casimir effect \cite{Fulling1976}, and the Schwinger effect \cite{Schwinger1951}. The resulting process can be seen as a scattering where it is possible to create particles of positive and negative norm and still fulfill the conservation of norm.

The existence of negative-norm modes $k_{u1},k_{u2}$ [$\omega_{u1},\omega_{u2}$] allow the creation of particles through an analog of the Hawking effect \cite{Jacobson1996}. In terms of the Bogoliubov transformation an outgoing mode is a linear superposition of two ingoing modes with opposite norm
\begin{equation}
\phi_\text{ur}^\text{out}=\alpha \phi_\text{ul}^\text{in}+\beta \phi_\text{u1}^\text{in}.
\end{equation}
From scalar product we obtain for the norm
\begin{equation}
1=|\alpha|^2-|\beta|^2,
\end{equation}
and if $|\beta|>0$ there is a norm amplification that can be thought of as creation of particles. If the vacuum state is amplified by an horizon, this is the analog of Hawking radiation \cite{Hawking1975}. In the dispersionless case, this radiation is in a thermal state described by
\begin{equation}
\left|\frac{\beta}{\alpha}\right|^2=\exp\left(-\frac{\hbar\omega}{k_B T}\right),
\label{relphase}
\end{equation}
where $\hbar$ is the reduced Planck constant and $k_B$ is the Boltzmann constant. For dispersive cases, the horizon becomes fuzzy in position ($z$ or $\tau$), which can be seen in numerical solutions of Hamilton equations in Eq. \eqref{hamsonic} and \eqref{hamoptic}. Therefore, it is better to characterize the horizon in the conjugated space ($k$ or $\omega$).

\begin{figure}
	\includegraphics[width=85mm]{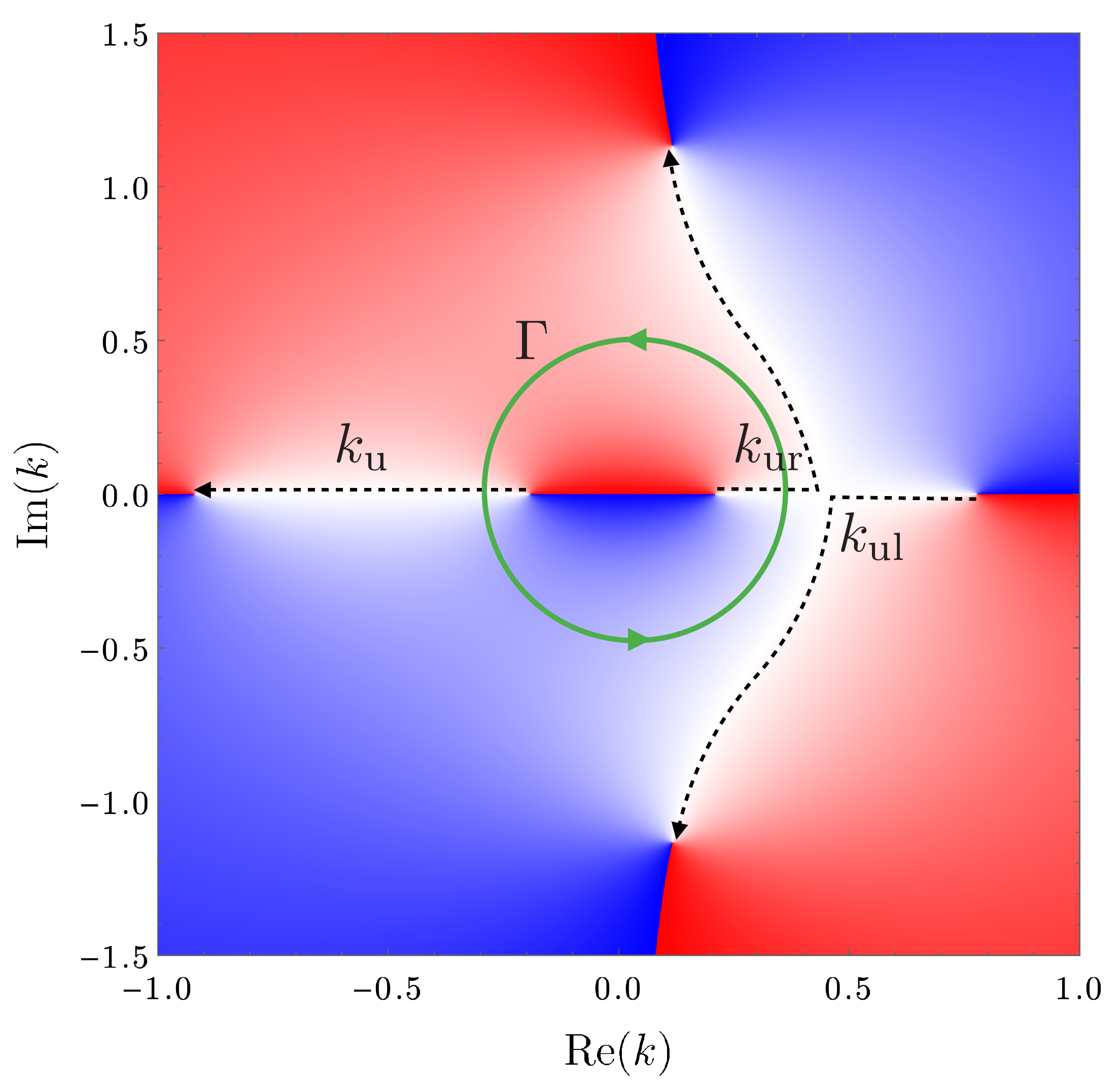}\vspace{3mm}
	\includegraphics[width=85mm]{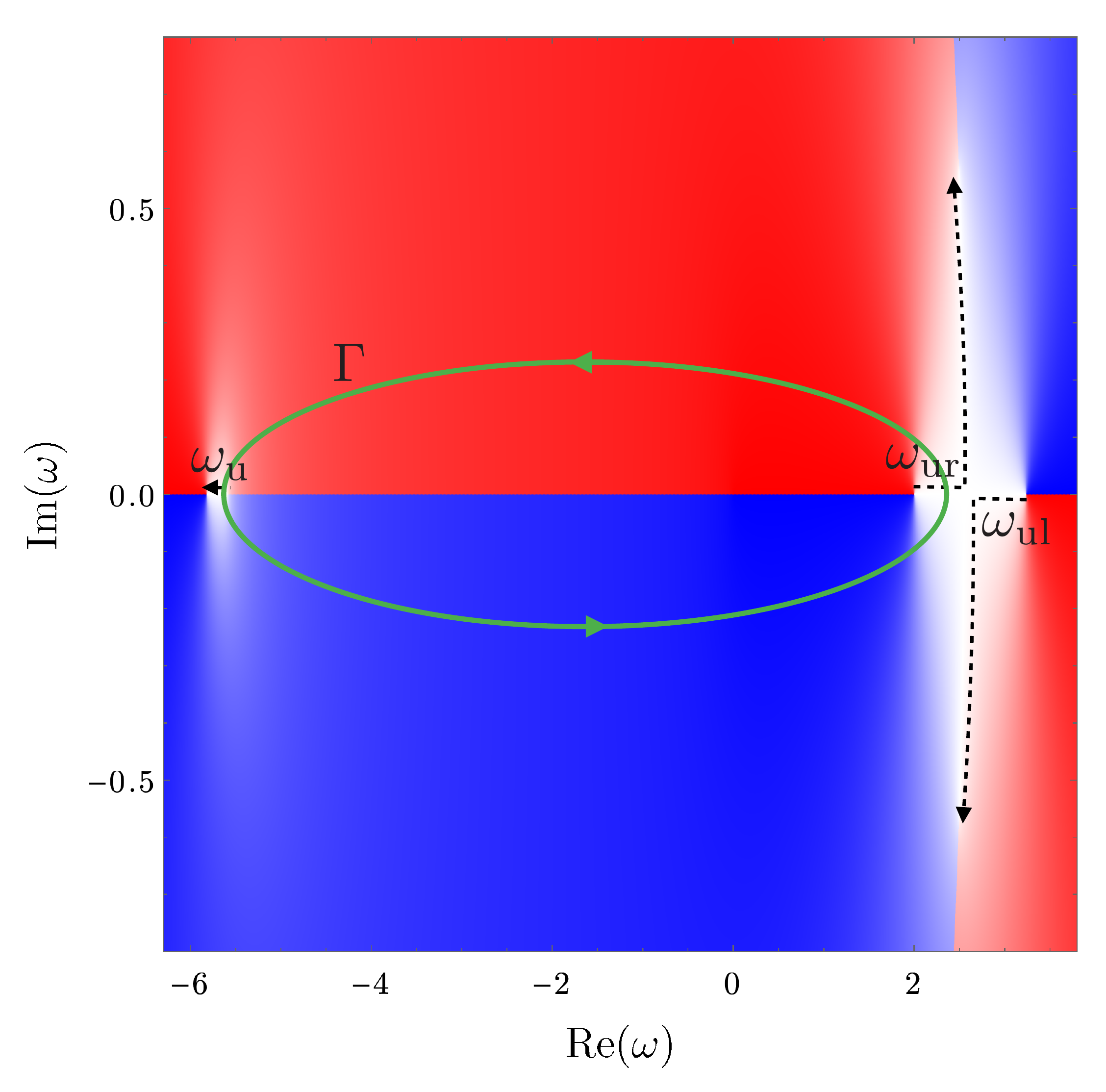}
	\caption{(Color online). Complex plane of $k$ for the sonic case (top) and of $\omega$ for the optical case (bottom). The value of Im$(z)$ [Im$(\tau)$] is color coded, negative in red, zero in white, and positive in blue. The real solutions are marked with black dashed lines. The green solid ellipses show the integration contour of Eqs. \eqref{intson} and \eqref{intopt}.}
	\label{complex}
\end{figure}

Leonhardt and Robertson \cite{Leonhardt2012} showed that the Hawking effect connects positive and negative norm modes in the complex plane. They developed an equation to calculate the Hawking temperature in sonic systems with dispersion. The relative phase $k$ of the Bogoliubov coefficients in the complex plane given by Eq. \eqref{relphase} is connected to a contour integral in position $z$, see Fig. \ref{complex}. This equation is
\begin{equation}
k_B T=\frac{i \hbar\omega}{\displaystyle\oint_\Gamma z(k)\dd k}.
\label{intson}
\end{equation}
The proof of this equation is done in the complex conjugated space using the saddle-point approximation and can be found in Appendix A of Ref. \cite{Leonhardt2012}.

In this work, we obtain an equivalent equation for the optical case by following the transformations in Section \ref{opticalanalog}. In this case the complex plane is in the laboratory frequency $\omega$ and the integration is in the delay time $\tau$
\begin{equation}
k_B T=\frac{i \hbar\omega}{\displaystyle\oint_\Gamma \tau(\omega)\dd \omega}.
\label{intopt}
\end{equation}
In Fig. \ref{complex} we can see the similarities and differences of the complex plane and the integration of the sonic and optics cases. The reason why these integrals are of the same form is because the topological structure of $\tau(\omega)$ is the same as $z(k)$ as shown in the figure, it is also why we connect the supersonic and subluminal dispersions \cite{Robson2019}.

In the next section, we will obtain the analytical Hawking temperature for general velocity profiles $z^{\pm 1/n}$ (sonic) and $\tau^{\pm 1/n}$ (optical). Two formulas for sonic analogs derived in Ref. \cite{Leonhardt2012} are included as particular cases, including the Schwarzschild spacetime.

Furthermore, we obtain an equation for a second-order approximation that can be used to obtain more detailed spectra in physical applications. In particular, we present results for a tanh and a sech${}^2$ profiles for the sonic case. These results are limited because, as we will see next, the analytical formula is given as an infinite series and is valid only when a convergence condition is fulfilled.

\section{Analytical Hawking temperature}
The solutions to integrals \eqref{intson} and \eqref{intopt} can be obtained in analytical form for profiles of the form $z^{\pm 1/n}$ or $\tau^{\pm 1/n}$ by using Cauchy's integral theorem. It is convenient to divide these integrals in growing and decaying profiles, i.e., if they grow or decay to the right side of the profile, as the method of solution depends on this (determined by the sign of the exponent: $+$ for growing and $-$ for decaying). In this section we present the derivation for a general profile $z^{\pm 1/n}$, but we work out explicitly six specific cases in the Appendix. Additionally, we only show the sonic case, as the analytic integrals are solved exactly the same way for the sonic and optical cases, unlike the numerical integrals we will discuss in Section \ref{sec:numerical}. In there, differences on the integration path become relevant.

\subsection{Growing profiles}
We can solve the integral for the sonic case given by Eq. \eqref{intson} for the velocity profiles expressed in the general form
\begin{equation}
u(z)=u_0\left(\frac{z}{a}\right)^{1/n},
\end{equation}
where $u_0, a$ are constants that define the scale of the profile and $n\in\mathbb{Z}$. For example, for $z$, $z^{1/2}$ and $z^{1/3}$ the results are shown in Table \ref{growing}. For comparison with the numerical solutions presented in the next Section, we also calculated the analytical formulas for $z^{1/4}$ and $z^{1/5}$, although we do not present the formulas here.

\begin{table}
	\centering
	\begin{tabular}{|l|l|}
		\hline
		$u(z)$ & $T/T_0$ \\
		\hline
		$z$ & $1$ \\
		$z^{1/2}$ & $\displaystyle\frac{c(k)}{v_g(0)}$ \\ 
		$z^{1/3}$ & $\displaystyle\frac{c^2(k)}{v_g^2(0)}\left(1+\frac{kc(k)v_g'(0)}{2v_g^2(0)}\right)^{-1}$\\
		\hline
	\end{tabular}
	\caption{Hawking temperature for growing profiles in the sonic case.}
	\label{growing}
\end{table}

Similarly, for optical profiles in the general form
\begin{equation}
\delta n (\tau) = \delta n_0 \left(\frac{\tau}{a}\right)^{1/n}
\end{equation}
For example, for the cases $\tau$, $\tau^{1/2}$ and $\tau^{1/3}$ the results are in Table \ref{growingop}. For simplicity, the results are expressed in terms of the velocities relatives to the perturbation: the relative phase-velocity $v_{pr}(\omega)=v_{g0}-c(\omega)$ and the relative group-velocity $v_{gr}(\omega)=v_{g0}-v_g(\omega)$.

\begin{table}
\centering
	\begin{tabular}{|l|l|}
		\hline
		$\delta n(\tau)$ & $T/T_0$                                        \\ \hline
		$\tau$           & 1                                              \\
		$\tau^{1/2}$     & $\displaystyle\frac{v_{pr}(\omega)}{v_{gr}(0)}$             \\
		$\tau^{1/3}$     & $\displaystyle\frac{v^2_{pr}(\omega)}{v^2_{gr}(0)}\left(-1+\frac{\omega v_{pr}(\omega)v_{gr}'(0)}{2v^2_{gr}(0)}\right)^{-1}$\\ \hline
	\end{tabular}
\caption{Hawking temperature for growing profiles for the optic case.}
\label{growingop}
\end{table}

\begin{figure}
	\flushleft
	\includegraphics[width=85mm]{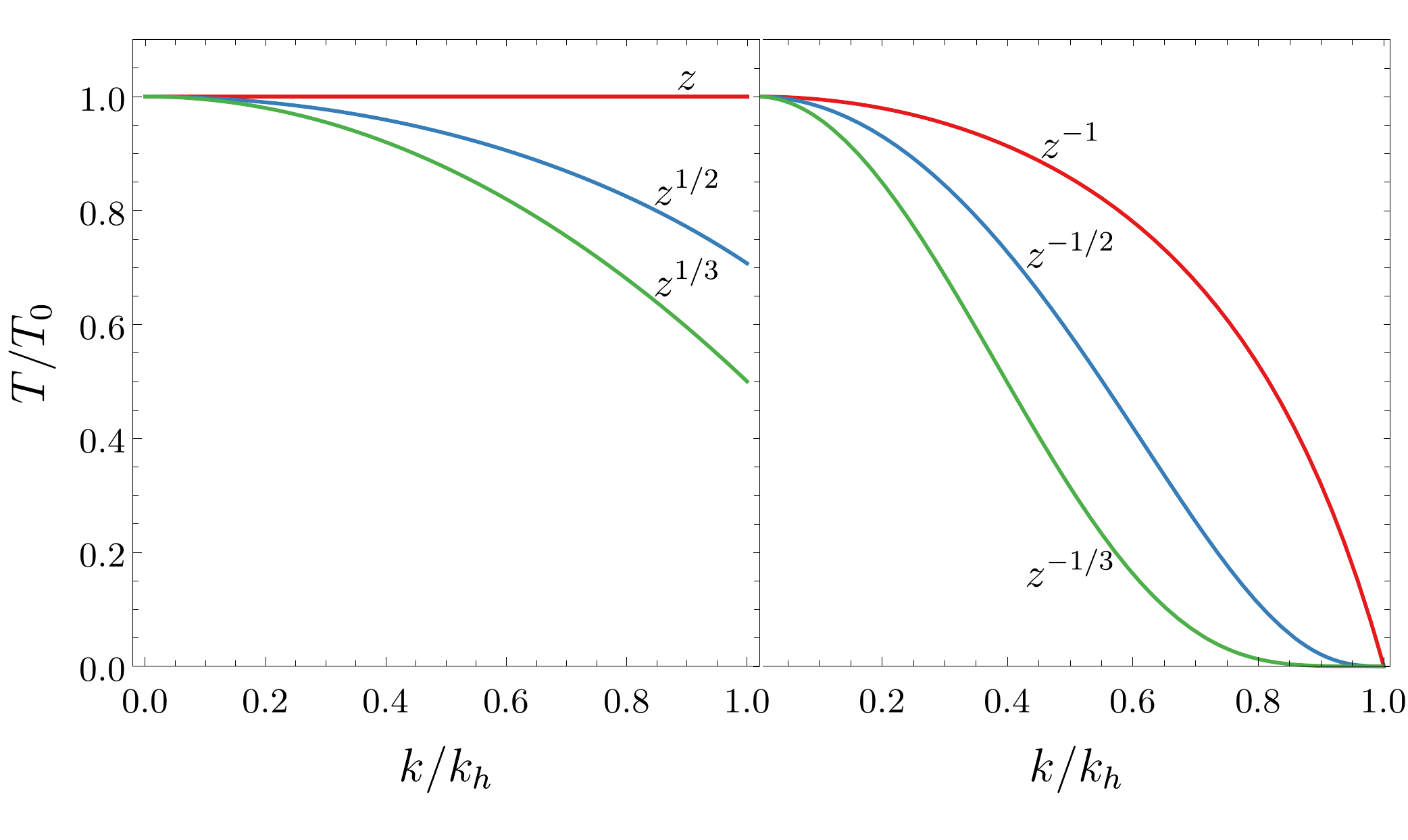}\vspace{-4mm}
	\includegraphics[width=85mm]{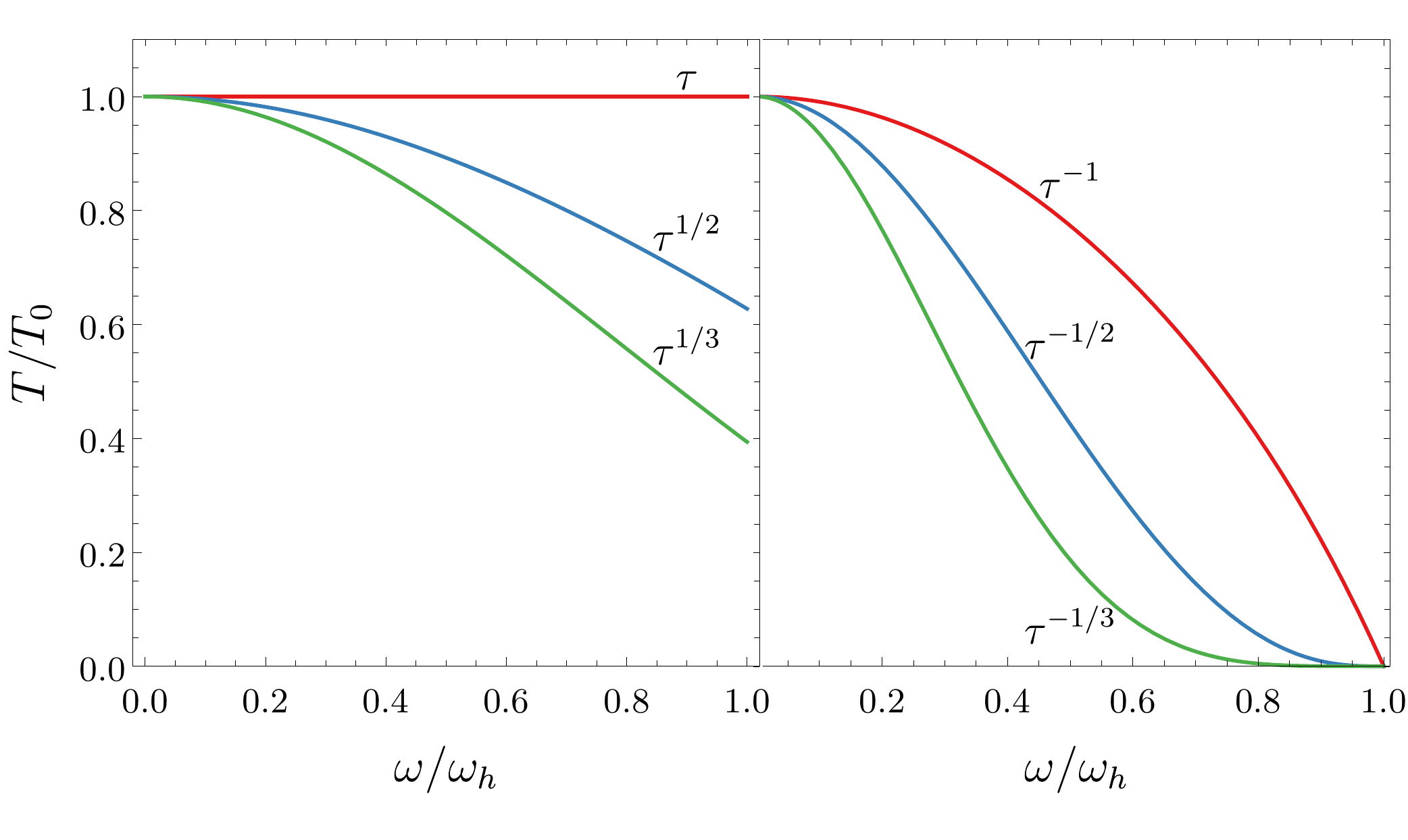}
	\caption{Hawking temperature for the sonic (top) and optical (bottom) cases. For both cases growing (left) and decaying (right) profiles are shown.}
	\label{Profiles}
\end{figure}

The resulting spectra are shown in terms of their effective Hawking temperature in Fig. \ref{Profiles}. A thermal emission would have a constant temperature throughout its spectra (as the $\propto z$ and $\propto\tau$ cases). The plots for the sonic analog are on top and for the optical analog at the bottom; the growing profiles are on the left-hand side and for decaying profiles on the right-hand side. The results are expressed in normalized units: to the thermal Hawking temperature and to the horizon wavenumber or frequency. As the velocity profiles diverge from 1, their spectra also depart from the thermal case as shown in Fig. \ref{Profiles}(left), this is due to the fact that the Hawking temperature depends on $c(k)$ or $v_{pr}(\tau)$ and the different velocity profiles. An attentive reader may be wondering why the Schwarzschild case $\propto z^{-1/2}$ is not thermal. The thermality of this profile is only for a dispersionless background, as the astrophysical spacetime is modeled in General Relativity. This was first shown by T. Jacobson \cite{Jacobson1991}. If the astrophysical spacetime has dispersion, e.g., at the Planck scale, then the Hawking spectrum would not be thermal even for the Schwarzschild case, it would be a good approximation for wavelengths shorter than the Planck length.

Now we will show the derivation of the analytic Hawking temperature for a growing velocity profile of the form
\begin{equation}
u(z)=u_0\left(\frac{z}{a}\right)^{1/n}.
\end{equation}
From the dispersion relation in Eq. \eqref{disprel} we have $u(z)=\omega/k-c(k)$, eliminating $u(z)$ we obtain
\begin{equation}
z(k)=\frac{au_0^nk^n}{(\omega-c(k)k)^n}.
\end{equation}
We develop the denominator in series around $k_0$ that fulfills $k_0c(k_0)-\omega=0$, i.e., the dispersion relation at $z = 0$, we only consider the first $n$ terms of the series as they are the only ones contributing to the integral
\begin{equation}
z(k)=\frac{au_0^nk^n}{\displaystyle\sum_{j=1}^{n}\frac{v_g^{j-1}(k_0)}{j!}(k-k_0)^j}.
\end{equation}
Using the residuals of the integral we have
\begin{align}
\frac{\hbar \omega}{k_B T}&=\displaystyle\frac{2 \pi a u_0^n}{v_g^n(k_0)(n-1)!}\times \nonumber\\
&\left.\partial_k^{n-1}\left(\frac{k^n}{\sum_{j=1}^{n}\frac{v_g^{(j-1)}(k_0)}{j!v_g(k_0)}(k-k_0)^{j-1}} \right)\right|_{k=k_0}.\label{eqgrowing}
\end{align}
At the phase horizon $c(k_0)=|u(z_0)|$ we can define the surface gravity as
\begin{equation}
\alpha=\frac{c^{n+1}(k_0)}{nau_0^n},
\end{equation}
and use it to simplify Eq. \eqref{eqgrowing}. The Hawking temperature is then
\begin{align}
T=&T_0\frac{v_g^n(k_0)}{c^n(k_0)}k_0 n!\times\nonumber\\
&\left[\left.\partial_k^{n-1}\left(\frac{k^n}{\sum_{j=1}^{n}\frac{v_g^{(j-1)}(k_0)}{j!v_g(k_0)}(k-k_0)^{j-1}} \right)\right|_{k=k_0}\right]^{-1},\label{eqgrowing2}
\end{align}
where $T_0$ is the usual Hawking temperature
\begin{equation}
T_0=\frac{\hbar\alpha}{2\pi k_B}.
\end{equation}

Equation \eqref{eqgrowing2} can be solved for different cases. For $n=1,2,3$, the solutions are given explicitly in Tables \ref{growing} and \ref{growingop} and worked out in the Appendix in Eqs. \eqref{solz1}, \eqref{solz2}, and \eqref{solz3}.

\subsection{Decaying profiles}
We can also solve the integrals in Eqs. \eqref{intson} and \eqref{intopt} for the velocity profiles in the general form
\begin{equation}
u(z)=u_0\left(\frac{a}{z}\right)^{1/n},\quad
u(\tau)=u_0\left(\frac{a}{\tau}\right)^{1/n},
\end{equation}
where again the constants define the scale of the profiles. For example, solving the integral of Eq. \eqref{intson}, the Hawking temperature for the first three profiles $z^{-1}$, $z^{-1/2}$ and $z^{-1/3}$ are shown in Table \ref{decaying}. Similarly, the integral of Eq. \eqref{intopt} for optical profiles is solved for  profiles $\tau^{-1}$, $\tau^{-1/2}$ and $\tau^{-1/3}$ and are shown in Table \ref{decayingop}. Again we can solve further profiles $z^{-1/4}$ and $z^{-1/5}$, but we do not present the results here.
\begin{table}
	\centering
	\begin{tabular}{|l|l|}
		\hline
		$u(z)$ & $T/T_0$ \\
		\hline
		$z^{-1}$ & $\displaystyle\frac{v_g(k)}{c(k)}$ \\
		$z^{-1/2}$ & $\displaystyle\frac{v_g^2(k)}{c^2(k)}\left(1-\frac{k v_g'(k)}{2v_g(k)}\right)^{-1}$ \\ 
		$z^{-1/3}$ & $\displaystyle\frac{v_g^3(k)}{c^3(k)}\left(1-\frac{3k v_g'{}^2(k)}{2v_g^2(k)}\right.$ \\
		 & $\displaystyle\qquad\left.+\frac{k^2}{6v_g^2(k)}[3v_g'{}^2(k)- v_g(k)v_g''(k)]\right)^{-1}$\\
		\hline
	\end{tabular}
	\caption{Hawking temperature for decaying profiles for the sonic case.}
	\label{decaying}
\end{table}

\begin{table}
	\begin{tabular}{|l|l|}
		\hline
		$\delta n(\tau)$ & $T/T_0$                                                                                                                                                                                                                 \\ \hline
		$\tau^{-1}$      & $\displaystyle \frac{v_{gr}(\omega)}{v_{pr}(\omega)}$                                                                                                                                                                                 \\
		$\tau^{-1/2}$    & $\displaystyle\frac{v^2_{gr}(\omega)}{v^2_{pr}(\omega)} \left(  1-\frac{\omega v_{gr}'(\omega)}{2v_{gr}(\omega)}   \right)^{-1}$                                                                                           \\
		$\tau^{-1/3}$    & $\displaystyle\frac{v^3_{gr}(\omega)}{v^3_{pr}(\omega)}\left(1-\frac{3\omega v'_{gr}(\omega)}{2v_{gr}(\omega)} \right.$\\
		&$\displaystyle\left.\quad +\frac{\omega^2}{6v_{gr}^2(\omega)}[ 3v'_{gr}{}^2(\omega)-v_{gr}(\omega)v''_{gr}(\omega)]    \right)^{-1}$\\ \hline
	\end{tabular}
\caption{Hawking temperature for decaying profiles in the optic analog.}
\label{decayingop}
\end{table}

The plots in Fig. \ref{Profiles}(right) show the results for the decaying profiles. The red line is for $z^{-1}$ and $\tau^{-1}$, the blue line for $z^{-1/2}$ and $\tau^{-1/2}$, and the green line for $z^{-1/3}$ and $\tau^{-1/3}$. It is clear that for these cases there is no thermal spectrum. Still, as the velocity profile gets further away from 1, the spectrum becomes less thermal and, in fact, for these type of profiles is always zero at the horizon. This fact results from the ratio of the phase-velocity and the group-velocity at the horizon.

To obtain the analytical solution for a decaying profile of the form
\begin{equation}
u(z)=-u_0\left(\frac{a}{z}\right)^{1/n},
\end{equation}
with $n\in\mathbb{Z}$, we solve the dispersion relation as
\begin{equation}
z=\frac{au_0^nk^n}{(kc(k)-\omega)^n}.
\end{equation}
We expand the denominator around $k_\infty$, the solution of the dispersion relation for $z\rightarrow\infty$, keeping only the first $n$ terms, as they will be the nonzero ones in the integral
\begin{equation}
kc(k)-\omega=\sum_{j=1}^{n}\frac{1}{j!}v_g^{(j-1)}(k_\infty)(k-k_\infty)^j.
\end{equation}
This solution has $n$ poles, however, only the simple pole at $k_\infty$ contributes to the integral, using Cauchy's residue theorem, we obtain
\begin{align}
\frac{\hbar\omega}{k_B T}=&\frac{2\pi}{(n-1)!}\frac{u_0^n}{v_g^n(k\infty)}\partial_k^{n-1}\left[ k^n\phantom{\frac{1}{1}}\right.\nonumber\\
&\left.\left.\times\left(\sum_{j=1}^n\frac{1}{j!}\frac{v_g^{(j-1)}(k_\infty)}{v_g(k_\infty)}(k-k_\infty)^{j-1}\right)^{-1}\right]^n\right|_{k=k_\infty}.
\end{align}
We calculate the surface gravity from the velocity profile at the phase horizon
\begin{equation}
\alpha=\left.\partial_z u(z)\right|_{z_\infty}=\frac{c^{n+1}(k_\infty)}{nau_0^n}.
\end{equation}
With this, we can simplify the Hawking temperature to
\begin{align}
T=&T_0\frac{v_g^n(k)}{c^n(k)}n!k \left[\partial_k^{n-1}k^n\phantom{\frac{1}{1}}\right.\nonumber\\
&\left.\left.\times\left(\sum_{j=1}^n\frac{1}{j!}\frac{v_g^{(j-1)}(k_\infty)}{v_g(k_\infty)}(k-k_\infty)^{j-1}\right)^{-1}\right]^n\right|_{k=k_\infty}.
\end{align}

\subsection{Second-order approximation}
Let us consider now a second-order velocity profile in the sonic case, i.e.,
\begin{equation}
u(z)=\alpha_1 z+\alpha_2 z^2.
\end{equation}
The dispersion relation \eqref{disprel} together with the binomial theorem leads to
\begin{equation}
z(k)=-\frac{\alpha_1}{2\alpha_2}\pm 2\sum_{m=0}^{\infty}a(m)\left(\frac{\omega}{k}-c(k)\right)^m.
\end{equation}
where
\begin{equation}
a(m)=\frac{(-1)^{m-1}}{2^m}\frac{(2m-3)!!}{m!(m-1)!}\frac{(4\alpha_2)^{m-1}}{\alpha_1^{2m-1}}.
\label{coeff}
\end{equation}

In this case, the Hawking temperature cannot be expressed as an analytical function because the integral in Eq. \eqref{intson} can only be solved as an infinite series. The result is
\begin{equation}
	T=\frac{\hbar \omega}{4\pi k_B}\left[\sum_{m=1}^\infty a(m)
	\partial_k^{m-1}\left(\omega-kc(k)\right)^m\big|_{k=0} \right]^{-1},\label{ts2}
\end{equation}

The necessary condition for the convergence of the series \eqref{ts2} is
\begin{equation}
\displaystyle\lim_{m\rightarrow\infty}\left|\frac{4\alpha_2(2m-1)}{\alpha_1^2(m^2+m)}\frac{\partial_k^m [\omega-kc(k)]^{m+1}}{\partial_k^{m-1}[\omega-kc(k)]^{m}}\right|<1.
\label{ccson}
\end{equation}

For the optical case the formula is similar, with the usual changes
\begin{equation}
T=\frac{\hbar \omega'}{4\pi k_B} \left[\sum_{m=1}^\infty a(m)\partial_\omega^{j-1}\left.\left( \omega'-\frac{\omega v_{pr}(\omega)}{c}\right)^j\right|_{\omega=0}\right]^{-1}.\label{2ndopt}
\end{equation}
This result is valid only when the following condition is fulfilled
\begin{equation}
	\displaystyle\lim_{m\rightarrow\infty}\left|\frac{4\alpha_2(2m-1)}{\alpha_1^2(m^2+m)}\frac{\partial_\omega^m \left( \omega'-\frac{\omega v_{pr}(\omega)}{c}\right)^{m+1}}{\partial_\omega^{m-1}\left( \omega'-\frac{\omega v_{pr}(\omega)}{c}\right)^{m}}\right|<1.
	\label{ccopt}
\end{equation}

\begin{figure}
	\flushleft
	\includegraphics[width=85mm]{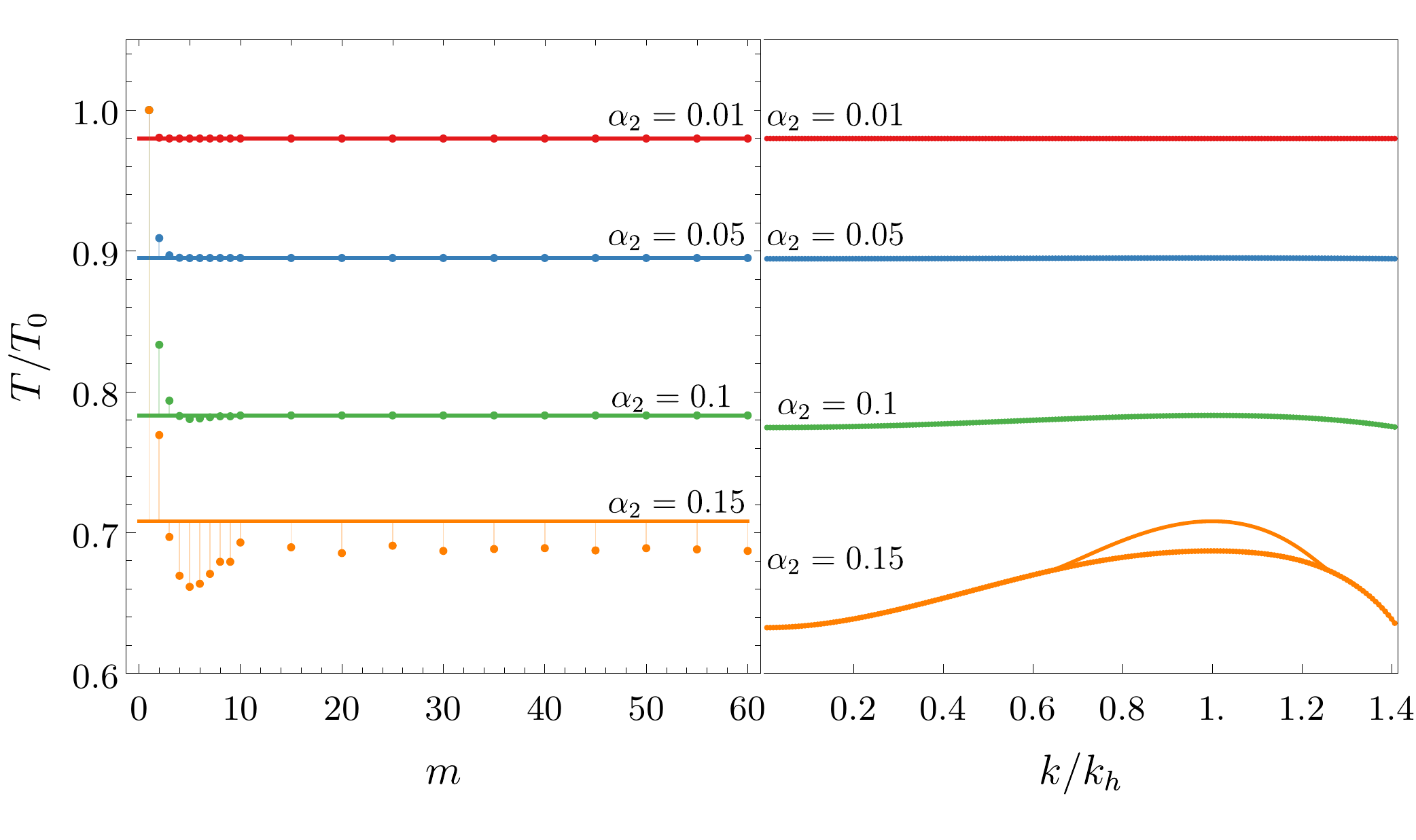}
	\caption{Hawking temperature for the second-order profile $\alpha_1 z+\alpha_2 z^2$ with $\alpha_1=1$, several values of $\alpha_2$, evaluated at the group-velocity horizon $k_h$, and with $m$ terms of the series (left). Hawking spectrum for the second-order profile with 60 terms (right). The lines indicate the numerical solution.}
	\label{con2nd}
\end{figure}

The series can be calculated analytically via symbolical software, we calculated up to 60 terms of the series. In Fig. \ref{con2nd}(left), a study of the convergence is shown at the group-velocity horizon, the most problematic area. If the value in Eq. \eqref{ccson} is much smaller than 1, the convergence is fast, however, if it is closer to 1 the series (points) converges to a different value than the numerical solution (line). Also note that the first term of all series is 1 as expected to match the linear approximation. In Fig. \ref{con2nd}(right) we studied the full spectra for several values of $\alpha_1, \alpha_2$, we see that the series does not match the numerical value for large $\alpha_2$ around the group-velocity horizon $k_h$. Besides, if $\alpha_2\rightarrow 0$, the spectrum becomes thermal as expected, but if this value increases, for a while the thermality is kept (constant $T/T_0$) but the temperature decreases ($T/T_0<1$). For even higher values thermality is finally lost, with $k_h$ dominating the spectrum.

\subsection{Second-order approximation to physical profiles}

In the analytical studies of Hawking radiation a linear approximation is usually taken, i.e., the spectrum is calculated considering that the velocity profile is linear $\propto z$ \cite{Unruh1995,Corley1999,Leonhardt2007} or $\propto \tau$ \cite{Philbin2008,Drori2019} and from there the famous formula for the Hawking temperature in analog systems is obtained
\begin{equation}
T=\frac{\hbar}{2\pi k_B} \left. \frac{\dd u}{\dd z}\right|_h,\qquad T= \frac{\hbar}{2\pi k_B}\left.\frac{1}{\delta n} \frac{\dd \delta n}{\dd \tau}\right|_h.
\end{equation}

In this work using the analytical formula in terms of the integral in the complex plane, we were able to obtain the Hawking temperature considering a second-order approximation given by Eqs. \eqref{ts2} and \eqref{2ndopt}. We can apply these formulas in the convergence region for the most used velocity profiles tanh and sech${}^2$ and obtain the difference with the linear profile. These velocity profiles together with the linear approximation form the base model for analog Hawking radiation.

\begin{figure}
	\flushleft
	\includegraphics[width=85mm]{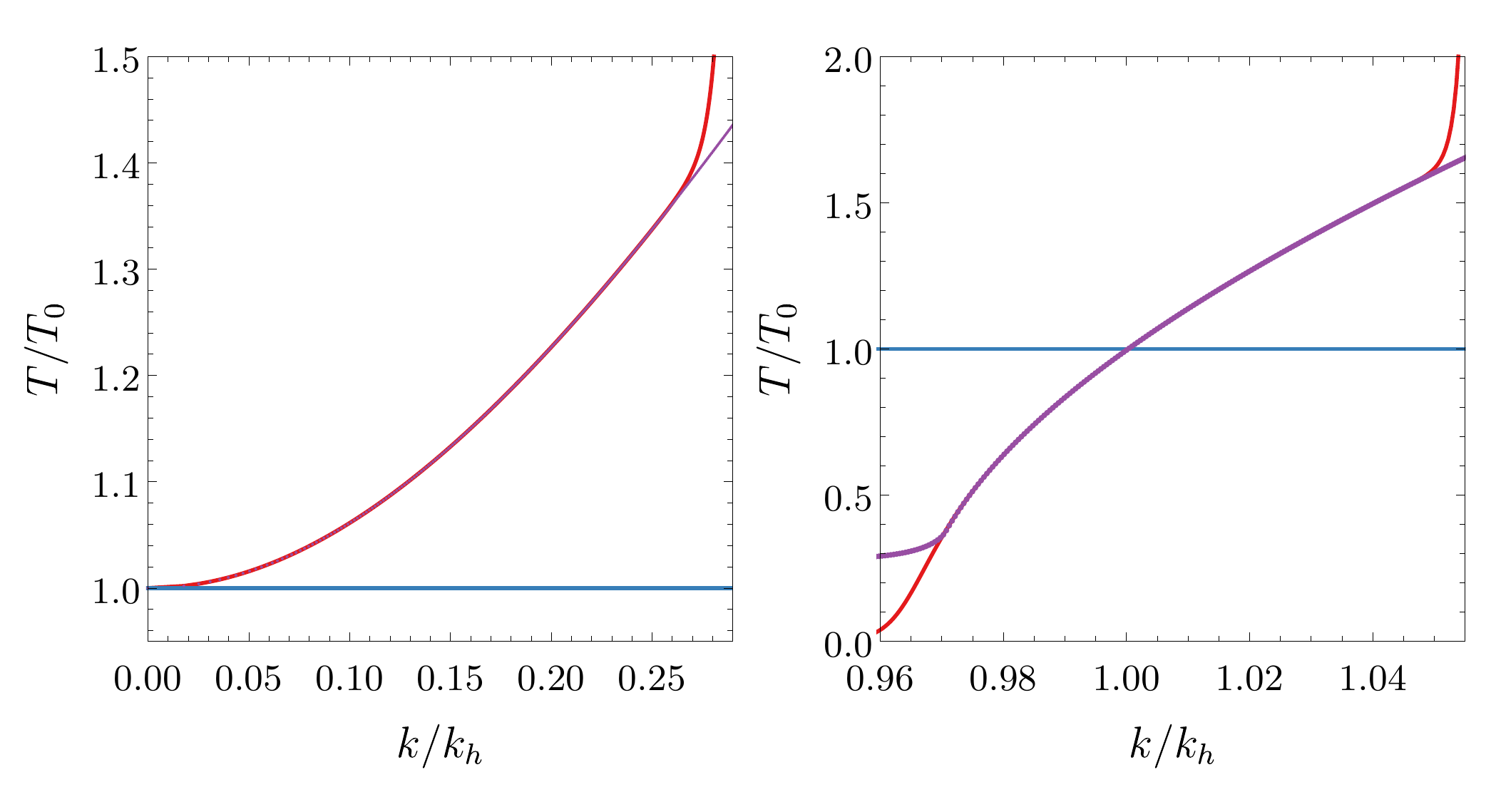}
	\caption{Hawking temperature for the tanh profile around $z=0$ and the sech${}^2$ around the zero curvature point. We show the analytical result for the first- (blue) and second-order (red) approximations and the numerical result (purple points).}
	\label{tanh}
\end{figure}

Finally, we compared our results with numerical solutions and obtained a good match as long as the series converges. This results are shown in Fig. \ref{tanh}. As expected, the spectra are thermal for low wavenumbers.

\section{Numerical Hawking temperature}\label{sec:numerical}
To obtain more information of the Hawking temperature from Eqs. \eqref{intson} and \eqref{intopt}, we can also perform numerical integrals to calculate the spectra. The numerical integration is performed in Cartesian coordinates, the trajectory is an ellipse that contains the origin and such that it never crosses any branch cuts in the complex plane, similar to the green contours shown in Fig. \ref{complex}.

\begin{figure}
	\flushleft
	\includegraphics[width=85mm]{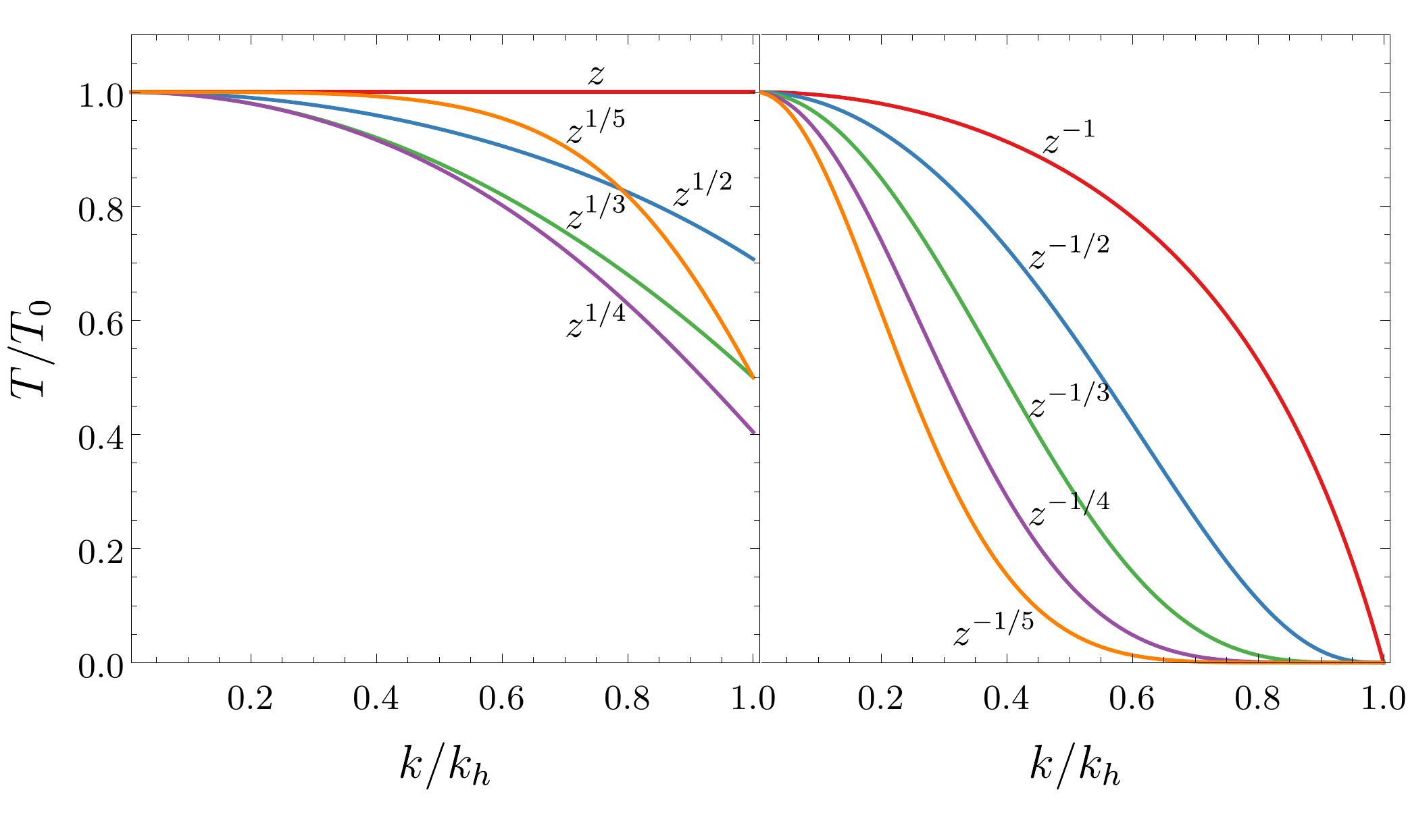}
	\caption{Hawking temperature for the sonic case obtained by numerical integration for comparison with the analytical results shown in Fig. \ref{Profiles}.}
	\label{num}
\end{figure}

For the cases that can be solved analytically, we confirmed that the numerical results were all the same up to the numerical error of $\sim\hspace{-1mm} 10^{-15}$. For example, we obtained the same spectra as in Fig. \ref{Profiles} that we present in Fig. \ref{num} for the sonic case. Additionally, we obtained the temperature for further velocity profiles $\pm 1/4$ and $\pm 1/5$. For the decaying cases we obtained that the dispersion is more important with higher roots, continuing the trend seen in the analytical cases. However, for the growing profiles the results for $1/5$ were unexpected, as they were closer to the thermal case ($T=T_0$) than expected. We confirmed these results by solving the analytical case for all these cases too. Although these formulas are not shown explicitly here, they match the numerical results in Fig. \ref{num}.

\section{Conclusions}\label{conclusions}
In this work we extended the analytical theory given by Leonhardt and Robertson \cite{Leonhardt2012} from the sonic to the optical case. We solved the resulting integral analytically for new velocity profiles and found the conditions for thermality, in particular, all of them are thermal for low wavenumbers (or frequencies in the optical case) as expected. Moreover, this formula can also be solved numerically, these evaluation match our analytical results. These numerical studies show us some unusual behavior for growing profiles that we then confirmed with further analytical studies.

For the optical case we found a similar formula for the Hawking temperature given in Eq. \eqref{intopt}. The solutions of the optical case in the complex plane are topologically equivalent to those in the sonic one. This leads to similar equations with a convenient choice of parameters and functions to describe the optical system. We solved the integral for growing and decaying perturbations analytically and numerically with excellent agreement.

We also obtained the analytical formula for a second-order velocity profile, although in this case the result is given as an infinite series. We were able to obtain the first 60 terms of the series and studied its convergence properties. The series converges when it fulfills conditions \eqref{ccson} or \eqref{ccopt} and its result matches the numerical solution.

One way to keep track that our analytical method is working correctly is that the real part of the integral in Eqs. \ref{intson} and \ref{intopt} is zero. If this is not the case, one of the main issues in the numerical integrals was the appropriate contour, once chosen correctly the error was of the order of $\sim\hspace{-1mm}10^{-15}$.

In analog theories the quantity that gives the order of the dispersion are the constants $k_0$ for the sonic and $\Omega_0$ for the optical cases. Those constants are related to the microscopic structure of the dispersive material that is perceived as a dispersion for large energies. Something similar could happen in real spacetime. If spacetime is dispersive, most likely its microscopic structure would be given by the Planck length and the dispersive effects could be felt for modes of comparable wavelengths. This would mean that even for the Schwarzschild spacetime the Hawking radiation would not be thermal for all frequencies, just to low frequencies compared with the Planck energy. This is because for a $z^{-1/2}$ profile with the dispersion, the Hawking temperature starts as the expected Hawking temperature $T_0$ and then it decreases \cite{Jacobson1991}.

\begin{acknowledgments}
AMR would like to acknowledge the financial support of Conacyt (Mexico) through scholarship 786398. The authors would like to thank the support from Fondo SEP-Cinvestav 602018.
\end{acknowledgments}

\appendix*
\section{Analytic spectra calculations}\label{appendix}
In this Appendix we will work out the method to obtain the analytical Hawking spectrum for several velocity profiles. We will formulate the sonic case, but the optical one is analogous.

\subsection{Profile $z$}
A linear velocity profile is
\begin{equation}
u(z)=u_0\frac{z}{a}=\alpha z.
\end{equation}
From the dispersion relation in Eq. \eqref{disprel} we have $u(z)=\omega/k-c(k)$, eliminating $u(z)$ we obtain
\begin{equation}
z(k)=\frac{\omega}{\alpha k}-\frac{c(k)}{\alpha}.
\end{equation}
Considering that $c(k)$ is an analytic function, the contour integral for the effective temperature \eqref{intson} is then
\begin{equation}
\int_\Gamma z(k)\dd k=\int_\Gamma\left(\frac{\omega}{\alpha k}-\frac{c(k)}{\alpha}\right)\dd k=
\frac{\omega}{\alpha}\int_\Gamma\frac{\dd k}{k}.
\end{equation}
Finally, the Hawking temperature is
\begin{equation}
T=\frac{\hbar\alpha}{2\pi k_B}\equiv T_0, \label{solz1}
\end{equation}
this is, the spectrum is thermal even with dispersion.

\subsection{Profile $z^{1/2}$}
For a square-root velocity profile
\begin{equation}
u(z)=u_0\left(\frac{z}{a}\right)^{1/2},
\end{equation}
and the dispersion relation \eqref{disprel} we have
\begin{equation}
z(k)=\frac{a}{(u_0 k)^2}(\omega-kc(k))^2.
\end{equation}
This function has a simple pole in $k=0$, by Cauchy's integral theorem we can integrate
\begin{equation}
\frac{\hbar\omega}{k_B T}=-\frac{4\pi \omega a v_g(0)}{u_0^2}.
\end{equation}
We also have $u(0)=0$, then from the dispersion relation we define $k_0$ as the solution of the dispersion relation that moves to the left $k_\text{ul}$ and from there $k_0c(k_0)=\omega$. Also, in the phase horizon $z_0$, the magnitude of the phase velocity is the same as the fluid velocity, i.e.,
\begin{equation}
c(k_0)=|u(z_0)|=u_0\left(\frac{z_0}{a}\right)^{1/2}.
\end{equation}
We can define $\alpha$ as the surface gravity of the system
\begin{equation}
\alpha= \left. \partial_z u(z)\right|_{z_0}=-\frac{u_0}{(az_0)^{1/2}}.
\end{equation}
From the last two equations we can obtain a relation between $a$ and $\alpha$ to simplify the effective temperature to
\begin{equation}
T=T_0 \frac{c(k_0)}{v_g(0)},\label{solz2}
\end{equation}
This is the value presented in Table \ref{growing}, where $k_0\rightarrow k$, as they refer to the unperturbed values of the dispersion relation.

\subsection{Profile $z^{1/3}$}
For a cubic root velocity profile
\begin{equation}
u(z)=u_0\left(\frac{z}{a}\right)^{1/3},
\end{equation}
the dispersion relation \eqref{disprel} leads to
\begin{equation}
z(k)=\frac{a}{(u_0 k)^3}(\omega-kc(k))^3.
\end{equation}
This function has a simple pole in $k=0$, by Cauchy's integral theorem we have
\begin{equation}
\frac{\hbar\omega}{k_B T}=-\frac{6\pi \omega a v_g^2(0)}{u_0^3}\left(1+\frac{\omega v_g'(0)}{2v_g^2(0)}\right).
\end{equation}
Following the same steps as in the previous case with $c(k_0)=|u(z_0)|$, the surface gravity is
\begin{equation}
\alpha=-\frac{u_0}{3(az_0^2)^{1/3}}.
\end{equation}
Then we can simplify the effective temperature to
\begin{equation}
T=T_0 \frac{c^2(k_0)}{v_g^2(0)}\left(1+\frac{c(k_0)k_0v_g'(0)}{2v_g^2(0)}\right)^{-1},\label{solz3}
\end{equation}
as presented in Table \ref{growing}.

\subsection{Profile $z^{-1}$}
For an inverse velocity profile
\begin{equation}
u(z)=-u_0\frac{a}{z},
\end{equation}
and from the dispersion relation we get
\begin{equation}
z(k)=\frac{au_0k}{kc(k)-\omega}.
\end{equation}
The only pole is in $z\rightarrow\infty$, where the velocity of the medium is zero $\omega=c(k_\infty)k_\infty$. Then, the integral is
\begin{equation}
\frac{\hbar\omega}{k_B T}=2\pi\frac{au_0k_\infty}{v_g(k_\infty)},
\end{equation}
and using the dispersion at infinity, we can calculate the surface gravity as
\begin{equation}
\alpha=\left.\partial_z u(z)\right|_{z_\infty}=\frac{c^2(k_\infty)}{au_0},
\end{equation}
and then simplify the spectrum to
\begin{equation}
T=T_0 \frac{v_g(k_\infty)}{c(k_\infty)}.\label{appm1}
\end{equation}

\subsection{Profile $z^{-1/2}$}
For the Schwarzschild profile or inverse square-root velocity profile
\begin{equation}
u(z)=-u_0\left(\frac{a}{z}\right)^{1/2},
\end{equation}
and from the dispersion relation we get
\begin{equation}
z(k)=\frac{au_0^2k^2}{(\omega-kc(k))^2}.
\end{equation}
This function has two poles, but only one contributes to the integral at $z\rightarrow\infty$, which results in
\begin{equation}
\frac{\hbar\omega}{k_B T}=\frac{au_0^2}{v_g^3(k_\infty)}(2v_g(k_\infty)k_\infty-v_g'(k_\infty)k_\infty^2).
\end{equation}
Again we calculate the surface gravity at the phase horizon
\begin{equation}
\alpha=\left.\partial_z u(z)\right|_{z_\infty}=\frac{c^3(k_\infty)}{2au_0^2},
\end{equation}
to simplify the Hawking spectrum to
\begin{equation}
T=T_0 \frac{v_g^2(k_\infty)}{c^2(k_\infty)}\left(1-\frac{v_g'(k_\infty)k_\infty}{2v_g(k_\infty)}\right)^{-1}.\label{appm2}
\end{equation}
This means that the velocity profile of Schwarzschild spacetime is not thermal in a dispersive medium.

\subsection{Profile $z^{-1/3}$}
For the inverse cubic-root velocity profile
\begin{equation}
u(z)=-u_0\left(\frac{a}{z}\right)^{1/3},
\end{equation}
and from the dispersion relation we get
\begin{equation}
z(k)=\frac{au_0^3}{(\omega/k-c(k))^3}.
\end{equation}
This function has three poles, but again the one that contributes to the integral is at $z\rightarrow\infty$, the result is
\begin{align}
\frac{\hbar\omega}{k_B T}=&2\pi\frac{3k_\infty au_0^3}
{v_g^3(k_\infty)}\left[1+\frac{k_\infty^2 v_g^2(k_\infty)}{2v_g(k_\infty)}\right.\nonumber\\ &\left.-\frac{k_\infty}{6v_g^2(k_\infty)}(9v_g'(k_\infty)+k_\infty v_g''(k_\infty))\right].
\end{align}
The surface gravity at the phase horizon
\begin{equation}
\alpha=\left.\partial_z u(z)\right|_{z_\infty}=\frac{c^4(k_\infty)}{3au_0^3},
\end{equation}
can be used to simplify the Hawking spectrum to
\begin{align}
T=T_0 &\frac{v_g^3(k)}{c^3(k)}\left(1-\frac{3v_g'{}^2(k)k}{2v_g^2(k)}\right. \nonumber\\
&\left.+\frac{k^2}{6v_g^2(k)}[3v_g'(k)^2- v_g''(k) v_g(k)]\right)^{-1}.\label{appm3}
\end{align}

\bibliography{references}
\end{document}